\documentclass[conference]{IEEEtran}
\IEEEoverridecommandlockouts
\usepackage{ifpdf}
\usepackage[dvips]{graphicx}
\graphicspath{{../eps/}}
\DeclareGraphicsExtensions{.eps}

\usepackage{mathbbol}
\usepackage{cite}
\usepackage[T1]{fontenc} % optional
\usepackage{algorithm}
\usepackage{algorithmic}
\usepackage[cmex10]{amsmath}
\usepackage{amsthm}
\usepackage{amssymb}
\usepackage{balance}
\usepackage{eqparbox}
\usepackage{textcomp}
\usepackage{ctable}
\usepackage{threeparttablex}
\usepackage{color}
\usepackage[cmintegrals]{newtxmath}

\newtheorem{theorem}{Theorem}

\usepackage[open,openlevel=1]{bookmark}
\usepackage{enumitem}

\allowdisplaybreaks

\newcommand{\bs}[1]{\boldsymbol{#1}}

\newcommand{\mb}[1]{\mathbf{#1}}
\newcommand{\mr}[1]{\mathrm{#1}}

\DeclareMathOperator*{\argmin}{arg\;min}

\makeatletter
\newcommand\fs@betterruled{%
  \def\@fs@cfont{\bfseries}\let\@fs@capt\floatc@ruled
  \def\@fs@pre{\vspace*{6pt}\hrule height.8pt depth0pt \kern2pt}%
  \def\@fs@post{\kern2pt\hrule\relax}%
  \def\@fs@mid{\kern2pt\hrule\kern2pt}%
  \let\@fs@iftopcapt\iftrue}
\floatstyle{betterruled}
\restylefloat{algorithm}
\makeatother

\newcommand{\bseq}{\begin{subequations}}
	\newcommand{\eseq}{\end{subequations}}
\newcommand{\baln}{\begin{align}}
	\newcommand{\ealn}{\end{align}}
\newcommand{\balnd}{\begin{aligned}}
	\newcommand{\ealnd}{\end{aligned}}
\newcommand{\beq}{\begin{equation}}
	\newcommand{\eeq}{\end{equation}}
\newcommand{\beqn}{\begin{eqnarray}}
	\newcommand{\eeqn}{\end{eqnarray}}
\newcommand{\beqno}{\begin{eqnarray*}}
	\newcommand{\eeqno}{\end{eqnarray*}}
\newcommand{\bma}{\begin{displaymath}}
	\newcommand{\ema}{\end{displaymath}}
\newcommand{\bnu}{\begin{enumerate}}
	\newcommand{\enu}{\end{enumerate}}
\newcommand{\bce}{\begin{center}}
	\newcommand{\ece}{\end{center}}
\newcommand{\btb}{\begin{tabular}}
	\newcommand{\etb}{\end{tabular}}
\newcommand{\ba}{\begin{array}}
	\newcommand{\ea}{\end{array}}

\begin{document}

	\bstctlcite{IEEEexample:BSTcontrol}
	\title{GEO Payload Power Minimization:  Joint Precoding and Beam Hopping Design}
	\author{\IEEEauthorblockN{Vu Nguyen Ha${}^{\dagger}$, Ti Ti Nguyen${}^{\ddagger}$, Eva Lagunas${}^{\dagger}$, Juan Carlos Merlano Duncan${}^{\dagger}$, and Symeon Chatzinotas${}^{\dagger}$\\
	\textit{$^{\dagger}$Interdisciplinary Centre for Security, Reliability and Trust (SnT), University of Luxembourg, Luxembourg} \\
		\textit{$^{\ddagger}$\'{E}cole de Technologie Sup\'{e}rieure (\'{E}TS), University of Qu\'{e}bec, Montreal, QC H3C 1K3, Canada}}
	}
	%\IEEEcompsocitemizethanks{Manuscript received January 25, 2019; revised October 14, 2019; accepted May 09, 2020. This work was supported in part by the National Sciences and Engineering Research Council of Canada under Grant RGPIN-2016-06401, and in part by Qu\'{e}bec's Merit Scholarship Program for Foreign Students from Minist\`{e}re de l'\'{E}ducation, de l'Enseignement Sup\'{e}rieur et de la Recherche du Qu\'{e}bec, FQRNT-PBEEE-2018, and in part by the Startup Fund from San Diego State University. The associate editor coordinating the review of this paper and approving it for publication was Antonia Tulino.}
	%\IEEEcompsocitemizethanks{Vu N. Ha and Jean-Fran\c{c}ois Frigon are with \'{E}cole Polytechnique de Montr\'{e}al, Poly-Grames Research Center, Montreal, Quebec, Canada, H3T 1J4 (e-mail: \{vu.ha-nguyen,j-f.frigon\}@polymtl.ca).}
	%\IEEEcompsocitemizethanks{Duy H. N. Nguyen is with Department of Electrical and Computer Engineering, San Diego State University, San Diego, CA, USA 92182 (e-mail: duy.nguyen@sdsu.edu).}
	%}
	
%	\markboth{IEEE Transactions on Wireless Communications}{HA \MakeLowercase{\textit{et al.}}: Joint Precoding and Beam Hoping Design for Operating Cost Minimization in GEO Satellite Communication Networks}
	
	%\renewcommand{\baselinestretch}{1.45}

	%\thispagestyle{empty}
	%\setlength\arraycolsep{2pt}
	%\setlength\arraycolsep{2pt}
	\maketitle

		\begin{abstract}
			This paper aims to determine linear precoding (LP) vectors, beam hopping (BH), and discrete DVB-S2X transmission rates jointly for the GEO satellite communication systems to minimize the payload power consumption and satisfy ground users' demands within a time window. Regrading constraint on the maximum number of illuminated beams per time slot, the technical requirement is formulated as a sparse optimization problem in which the hardware-related beam illumination energy is modeled in a sparsity form of the LP vectors. To cope with this problem, the compressed sensing method is employed to transform the sparsity parts into the quadratic form of precoders. Then, an iterative window-based algorithm is developed to update the LP vectors sequentially to an efficient solution. Additionally, two other two-phase frameworks are also proposed for comparison purposes. In the first phase, these methods aim to determine the MODCOD transmission schemes for users to meet their demands by using a heuristic approach or DNN tool. In the second phase, the LP vectors of each time slot will be optimized separately based on the determined MODCOD schemes.
		\end{abstract}
		
%		\begin{IEEEkeywords}
%			Multibeam satellite, Precoding, Beam Hopping.
%	\end{IEEEkeywords}
	
	\vspace{-0.1cm}
%	\IEEEpeerreviewmaketitle
	
%	\maketitle
%	\IEEEdisplaynotcompsoctitleabstractindextext
%	\IEEEpeerreviewmaketitle
	
	\section{Introduction}
High-Throughput Satellite (HTS) systems have been identified as key solution to deliver ubiquitous,
high-quality connectivity globally. Furthermore, multi-beam HTS (MB-HTS) has dramatically improved the satellite system throughput by allocating limited radio resources uniformly across beams \cite{Sat_Survey_21}.
Recently, LP and BH have been considered as two promising technologies for MH-HTS. Particularly, BH strategies can illuminate a number of beams at each time slot (TS) to balance between the traffic flows going through the payload and operating cost efficiently \cite{Sat_Survey_21} while the LP technique is an effective processing tool for mitigating the inter-beam interference and improving the network performance significantly \cite{{Vazquez_WC16}}. The concept of using these two advanced techniques jointly have attached much attention from both industry, i.e., ESA \cite{Ginesi_WiSATS17,ESA_FlexPreDem} and academia \cite{Kibria_Globe19,Eva_Frontier21}.
However, in these works, the LP technique is only employed to increase the capacity of the satellite channels once the BH solution has been defined.

To the best of our knowledge, the joint design of BH, LP, and DVB-S2X-based MODCOD scheme selection for multi-beam GEO satellite communication systems has not been considered in the literature. This paper focuses on tackling this subject by minimizing the payload power consumption and supporting the ground users at their data traffic demand during a window consisting of number of TSs.
By explicitly considering the effect of illuminating each beam on the required power for hardware-related and transmission processes, we describe the total power consumption in a mixed sparsity and quadratic form of the LP vectors.
The technical requirement is then formulated as a sparse mixed-discrete optimization problem which is well-known as NP-hard.
To deal with this challenging problem, we
employ re-weighted quadratic-form relaxation method method to deal with the sparsity and relax the discrete rate function into a continuous form by utilizing Matlab fitting tool.
Afterward, three algorithms are presented to obtain the efficient LP and BH solutions. 
In particular, the first approach aims to jointly optimize all  transmission rates and LP vectors right at the TS one by alternatively optimizing precoders and updating sparsity-relaxing weights.
The other two are developed based on a two-phase solution approach.
In the first phase, the MODCOD schemes with various transmission rates are predetermined for all users to meet their demands by equally allocating the transmission rate over all TSs or employing the effective (Deep Neuron Network) DNN tool.
In the second phase, thanks to the MODCOD selected in the precious, the LP vectors of each TS are then optimized separately.  
Finally, numerical results are presented to demonstrate convergence as well as superior performance of the proposed algorithms.

\vspace{-0.2cm}

	\section{System Model and Problem Formulation} \label{sec:SMnPF}
	\begin{figure}[!t]
		\centering
		\includegraphics[width=60mm]{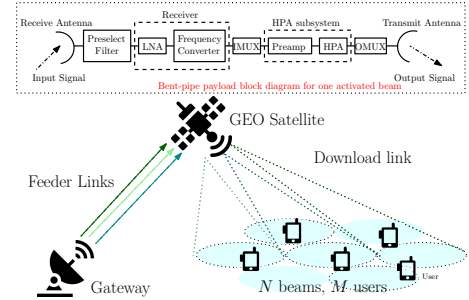}
		\vspace{-0.2cm}
		\caption{A multi-beam GEO satellite communication system.}
		\label{GEO_to_User}
			\vspace{-0.5cm}
	\end{figure}
	
	Consider a forward link of a broadband multibeam satellite system with $N$ beams serving $M$ ground users within a time window of $T$ TSs (Fig.~\ref{GEO_to_User}).
	Regarding user service demand, the amount of data stored in HTS system memory corresponding to $M$ users, denoted as $\bar{Q}_1, ... , \bar{Q}_M$ (bits), which should be delivered to $M$ users no later than TSs $\bar{T}_1,...,\bar{T}_M$, respectively.
	Denote $\mathcal{N}$ as the set of all beams.
	Thanks to LP and BH techniques, different subsets of $\mathcal{N}$ can be selected to serve a specific user in different TSs.
    In addition, if beam $n$ is assigned to serve user $m$ in TS $t$, an LP factor $w_{n,m}[t] \in \mathbb{C}$ is applied to the corresponding data symbol. 
    %Without loss of generality, we assume $\mathbb{E}\{|s_{m}[t]|\} = 1$.
	Certainly, $ w_{n,m}[t]  = 0$ implies that no service from beam $n$ to user $m$ within this TS.
	Let $h_{n,m}[t] \in \mathbb{C}$ be the channel coefficient due to beam $n$ and user $m$ in TS $t$. Denote $\mb{w}_m[t] = \left[w_{n,m}[t]\left|_{n = 1,...,N} \right.  \right]$ and $\mb{h}_{m}[t] = \left[h_{n,m}[t]\left|_{n=1,...,N }\right.   \right]$ ($\mb{w}_m[t], \mb{h}_{m}[t] \in \mathbb{C}^{N \times 1}$) as the LP and the channel vectors corresponding to user $m$ in TS $t$. 
	Then, received signal at user $m$ in TS $t$ can be described as
	$z_m[t]  = \sum_{j = 1}^{M} \mb{h}^H_{j,m}[t] \mb{w}_j[t] s_{j}[t] + \eta_m[t]$ based on which the corresponding SINR can be written as
	\beqn
	\Gamma_m[t] = {\left|\mb{h}^H_{m,m}[t] \mb{w}_m[t] \right|^2}/{(\sum \limits_{j \neq m} \left|\mb{h}^H_{j,m}[t] \mb{w}_j[t] \right|^2  + \sigma_m^2)}.
	\eeqn

	\vspace{-0.5cm}

\subsection{DVX-S2X MODCOD Schemes and Achievable Rate}
	Following to DVB-S2X standard \cite{ETSI_DVBS2X}, one MODCOD scheme out of $L$ candidates will be selected for the data transmission of each user within a TS. 
	Denote $\mathcal{R}_{\sf{DVB}}=\{0,R_1,...,R_L\}$ the set of zero value and $L$ data rates corresponding to $L$ MODCOD schemes.
	ETSI technical report in \cite{ETSI_DVBS2X}) has suggested different minimum target SINR according to each MODCOD scheme for a specific block-error-rate (BLER) result. 
	Let $\Omega = \{0,\bar{\gamma}_1,\bar{\gamma}_2,...,\bar{\gamma}_L\}$, where $0<\bar{\gamma}_1<...<\bar{\gamma}_L$, be the set of all target SINR corresponding to the transmission rates in $\mathcal{R}_{\sf{DVB}}$ for a specific pre-determined BLER target, e.g., $10^{-5}$ as in Table 20a-b-c in \cite{ETSI_DVBS2X}.
	
	Let $g_m[t] \in \Omega$ be the discrete variable presenting the MODCOD selection of user $m$ in TS $t$. Herein, $g_m[t] = \bar{\gamma}_l$ means the selection of MODCOD scheme $l$. Furthermore, $g_m[t]=0$ presses that no transmission for user $m$ is processed within that TS.
		Let us present a mapping function $f_{\sf{DVB}}(x): \Omega \mapsto \mathcal{R}_{\sf{DVB}}$ where
	$f_{\sf{DVB}}(0) = 0 \text{ and } f_{\sf{DVB}}(\bar{\gamma}_l) = R_l, \forall l$.
	Then, the data rate of user $m$ in TS $t$ can be estimated as
	$R_m[t] = \Delta_T BW f_{\sf{DVB}}(g_m[t])$ where $BW$ and $\Delta_t$ are the bandwidth and TS duration.
	In addition, the target SINR requirement can be stated as
	\beq \label{eq:modcod_cnt2}
	(C1): \quad \Gamma_m[t](\mb{w}_t) \geq g_m[t], \quad \forall (t,m).
	\eeq  
	
	\vspace{-0.4cm}
	
	\subsection{Beam Hopping and Payload Power Consumption} 
	Let $\mathcal{M}$ be the set of all users. Then, the transmission power of beam $n$ in TS $t$ can be expressed as
	\beq \label{eq:Pt}
	P_n[t] = \sum \limits_{m \in \mathcal{M}} w_{n,m}[t]^{\prime} w_{n,m}[t] = \sum \limits_{m \in \mathcal{M}} \mb{w}_{m}^H[t] \mb{E}_m \mb{w}_{m}[t].
	\eeq 
	where $\mb{E}_n$ is a diagonal matrix in $\mathbb{R}^{N \times N}$ with zero elements and one at the $n$-th position.
	It can be observed that beam $n$ is activated in TS $t$ if and only if $P_n[t] > 0$.
	Due to the payload limitation, we further assume that the GEO HTS can only illuminate at most $K_t$ beams in TS $t$, i.e.,
	\beq \label{eq:xPw}
	(C2): \quad \sum \limits_{n \in \mathcal{N}} \left\| P_n[t] \right\|_0 \leq K_t, \quad  \forall t.
	\eeq
    In addition, illuminating one beam for transmission requires the operation process of several hardware elements, such as, pre-select filter, low-noise amplifier, frequency converter, input/output multiplexers, pre-amplifier, and high-power amplifier, etc.. Taking into account all power consumption due to these components, illuminating one beam results in a constant hardware-related power \eqref{eq:Pt}.
	Denote this amount as $\rho_{\sf{hw}}$, the total beam illumination hardware-related power can be described as
	\beq
	P_{\sf{illu}} = \rho_{\sf{hw}} \sum \limits_{t \in\{1,...,T\}}\sum \limits_{n \in \mathcal{N}} \left\| P_n[t] \right\|_0.
	\eeq
	
		\vspace{-0.4cm}

	\subsection{Problem Formulation}
	%\begin{figure}[t!]
	%\centering
	%\includegraphics[width=90mm]{Diagram_solution_approach.eps}
	%\caption{Diagram of solution approach.}
	%\label{flow-chart-fig}
	%\end{figure}
	This work focuses on optimizing joint LP and BH design to minimize the payload power consumption while satisfying all users at their demands. This problem can be stated as
	\begin{subequations} \label{MIN_POWER_PROB}
		\begin{eqnarray} 
			\hspace{-0.8cm}&\underset{\mb{W},\mb{g}}{\min}& \hspace{-0.2cm}  \sum \limits_{\forall (n,t)} \!\!\! \left( P_n[t] + \rho_{\sf{hw}} \left\| P_n[t]\right\|_0\right)  \label{obj_func_SEE}\\
			\hspace{-0.8cm}&\text{s.t. }&  \hspace{-0.2cm}  \text{constraints $(C1)$, $(C2)$,} \nonumber \\
			\hspace{-0.8cm}&& \hspace{-0.2cm} (C3): \quad P_n[t] \leq \bar{P}^{\mr{GEO}}_{n}, \quad \forall (n,t), \label{cnt1} \\
			\hspace{-0.8cm}&& \hspace{-0.2cm} (C4): \quad \sum_{t \in \{1,...,\bar{T}_m\}}R_m[t]  \geq \bar{Q}_m, \quad \forall m, \label{cnt5} \\
			\hspace{-0.8cm}&& \hspace{-0.2cm} (C5): \quad	g_m[t]  \in \Omega , \quad \forall (m,t), \label{cnt6}
		\end{eqnarray}
	\end{subequations}
	where $\mb{W},\mb{g}$ stand for the matrix and vector representing all LP vectors and selected target SINRs, and $\bar{P}^{\mr{GEO}}_{n}$ is the maximum transmission power of beam $n$.
	Problem (\ref{MIN_POWER_PROB}) is a complicated long-term optimization
	problem.
	%The challenges for solving problem (\ref{MIN_POWER_PROB}) not only come from the sparse norm-$\ell_0$ term in the objective function and the coupling between binary and complex variables but also that the payload delivery depends on the resource allocation strategy in each time slot.
		\vspace{-0.1cm}
	
	\section{Window-based Optimization Solution} \label{sec:03}
	As can be observed, problem \eqref{MIN_POWER_PROB} is NP-hard due to the discrete variables, $g_{m}[t]$'s, and the $\ell_0$-norm terms in both objective function and constraint $(C2)$.
	In what follows, we first degrade the complexity level of solving this problem by relaxing the discrete variables to the continuous ones and approximating the sparsity term to the quadratic form of LP vectors.
	Then, an iterative algorithm is proposed to deal with the relaxed problem.
	
		\vspace{-0.2cm}
		
	\subsection{Problem Approximation} 
	\subsubsection{Discrete-to-Continuous Approximation}
	    	\begin{figure}[!t]
		\centering
		\includegraphics[width=65mm]{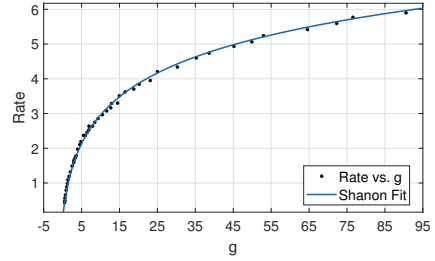}
		\vspace{-0.3cm}
		\caption{DVB-S2X rate vs. $g$ and Shanon fitting curve.}
		\label{DVB-S2X-fitting_fig}
		\vspace{-0.5cm}
	\end{figure}
	The discrete variables can be relaxed into the continuous ones by replacing the constraints given in $(C5)$ as 
	\beq \label{cnt:gm_rlx}
	(\tilde{C}5): \quad 0 \leq g_m[t] \leq \bar{\gamma}_L, \quad \forall (m,t).
	\eeq
	Adopting a practical data-fitting tool, the mapping function $f_{\sf{DVB}}(*)$ can be replaced by 
	\beq \label{eq:shanon_appr}
	f_{\sf{SN}}(g_m[t])=\min\left[ R_L,\log_2\left(1+g_m[t]/\xi_{\sf{fit}}\right)\right] ,
	\eeq
	where $\xi_{\sf{fit}}$ is the fitting parameter which represents the MODCOD loss in comparison to Shanon bound. 
	Exploiting the values of code rate and the corresponding SNR from Table 20a-b-c in \cite{ETSI_DVBS2X} and the Matlab fitting tool, the fitting function $f_{\sf{SN}}(g)$ is illustrated in Fig.~\ref{DVB-S2X-fitting_fig} where $\xi_{\sf{fit}} = 1.473$ and the root mean square error ($RMSE$) equal to $0.06884$. Exploiting this approximation, constraint $(C4)$ can be re-stated as
	\beq
	(\tilde{C}4): \quad \sum \limits_{t \in \{1,...,\bar{T}_m\}}  f_{\sf{SN}}(g_m[t]) \geq \bar{Q}_m/(\Delta_TBW), \forall m. \label{cnt4aa}
	\eeq
%	Next, we characterize the relation of $g_m[t]$'s and $\Gamma_m[t]$'s at the optimal point in the following theorem.
	\begin{theorem} \label{SINR_g_thr1} Let $(\mb{g}^{*},\mb{W}^{*})$ be the optimal solution of problem \ref{MIN_POWER_PROB}. Then, the constraint $(C1)$ must hold with equality, specifically,
	$ \Gamma_m[t](\mb{w}^*_t) = g_m^*[t], \forall (t,m)$.
	\end{theorem}
	\begin{IEEEproof} 
	Assume that at the optimal point $(\mb{g}^{*}$, $\mb{W}^{*})$, there is at least one couple $(\hat{t},\hat{m})$ that the corresponding constraint $(C1)$ holds with inequality, i.e., 
$\Gamma_{\hat{m}}[\hat{t}](\mb{w}^*_{\hat{t}}) > g_{\hat{m}}^*[\hat{t}]$.
		Then, let's define new LP set $\mb{W}^{\prime}$ that $\mb{w}_{\hat{m}}^{\prime}[\hat{t}] = \sqrt{{g_{\hat{m}}^*[\hat{t}]}/{\Gamma_{\hat{m}}[\hat{t}](\mb{w}^*_{\hat{t}})}} \mb{w}_{\hat{m}}^{*}[\hat{t}]$ and $\mb{w}_m^{\prime}[t] = \mb{w}_m^{*}[t]$ for all $(t.m) \neq (\hat{t},\hat{m})$. It is easy to see that  $(\mb{g}^{*},\mb{W}^{\prime})$ is also a feasible solution of  problem \eqref{MIN_POWER_PROB} but the objective function due to $(\mb{g}^{*},\mb{W}^{\prime})$ is less than that of $(\mb{g}^{*},\mb{W}^{*})$ which results in a contradiction. Therefore, the proof must follow.
	\end{IEEEproof}
Then, $(C1)$, $(\tilde{C}4)$, and $(\tilde{C}5)$ can be replaced by
\beq
(C6): \sum_{t \in \{1,...,\bar{T}_m\}}  f_{\sf{SN}}(\Gamma_{m}[t](\mb{w}_{t})) \geq \bar{Q}_m/(\Delta_TBW), \forall m. \label{cnt4bb}
\eeq

	\subsubsection{Sparsity Approximation}
	To deal with the sparsity issues, we employ the re-weighted $\ell_1$-norm minimization method \cite{Candes08}. 
	The method tends to utilize special weights on the sparsity related elements, which is updated iteratively in order to relax the $\ell_0$-norm terms to $\ell_1$-norm form as closed as possible.
	Employing this method,  $\left\| P_n[t]\right\|_0$ can approximate to
	\beq \label{eq:CSapprx}
	\left\| P_n[t]\right\|_0 \approx \psi_n^{(k)}[t] P_n[t],
	\eeq
	where $\psi_n[t]^{(k)}$ is the $\ell_1$-norm relaxing weight applied in iteration $k$ of the outer loop.
	Utilizing a small scalar $\varepsilon$, i.e., $\varepsilon \ll 1$, following \cite{Candes08}, one can determine $\psi_n[t]^{(k)}$ based on the value of $P_n[t]$ in iteration $k$, i.e., $P_n[t]^{(k)}$, as
	\beq 
	\psi_n^{(k)}[t] = \left( P_n^{(k)}[t]^2 + \varepsilon\right)^{-1/2}. \label{weight_factor3}
	\eeq
	
	\subsubsection{Relaxed Problem}
	The approximation results given in \eqref{eq:shanon_appr} and \eqref{eq:CSapprx} facilitate us to develop an efficient iterative algorithm to solve problem \eqref{MIN_POWER_PROB} by considering the following problem in each iteration.
	\begin{subequations} \label{MinPro_rlxed2}
		\begin{eqnarray} 
			\hspace{-0.8cm}&\underset{\mb{W}, \mb{g}}{\min}& \hspace{-0.2cm} \sum \limits_{\forall (t,n)}  \left( 1 + \rho_{\sf{hw}} \psi_n^{(k)}[t]\right) P_n[t]  \label{obj_rlxed}\\
			\hspace{-0.8cm}&\text{s. t. }&  \hspace{-0.2cm}   \text{$(C3)$, $(C6)$, and }  (\tilde{C}2): \sum \limits_{\forall n} \psi_n^{(k)}[t] P_n[t]  \leq K_t, \forall t, \label{cnt3aa}
		\end{eqnarray}
	\end{subequations}
	
	\vspace{-0.4cm}
	
	\subsection{Iterative Solution Approach}
	\subsubsection{Dual Problem}
Let $\beta_n^{(k)}[t] = \left( 1 + \rho_{\sf{hw}} \psi_n^{(k)}[t]\right)$, 
the Lagrangian of \eqref{MinPro_rlxed2} regarding to $(C6)$ can be described as
\beqn 
&&\hspace{-1.2cm} \mathcal{L}(\mb{W},\bs{\mu})  =  \sum \limits_{\forall (t,n)}  \beta_n^{(k)}[t] P_n[t]   \nonumber \\
&&\hspace{0.5cm}   + \sum \limits_{\forall m} \mu_m [  \dfrac{\bar{Q}_m}{\Delta_TBW}- \sum \limits_{t = 1}^{\bar{T}_m}  f_{\sf{SN}}\left( \Gamma_{m}[t](\mb{w}_{t})\right)  ], 
\eeqn
where $\mu_m \geq 0$ is the Lagrangian multiplier corresponding to user $m$.
Then, the dual problem of \eqref{MinPro_rlxed2} can be given as
\beq \label{dual-prob}
\underset{\bs{\mu}}{\max} \; \sf{g}(\bs{\mu}) \text{ s. t. } \mu_m \geq 0, \forall m,
\eeq
where $\sf{g}(\bs{\mu})$ is the dual function which is determined as
\beq \label{dual_func}
\sf{g}(\bs{\mu}) = \underset{\mb{W}}{\min} \; \mathcal{L}(\mb{W},\bs{\mu})  \text{ s. t. $(\tilde{C}2)$ and $(C3)$.}
\eeq
As can be seen, the strong duality holds if problem \eqref{MinPro_rlxed2} is feasible since there exists $\bs{\mu}$ that optimal value of \eqref{MinPro_rlxed2} equal to that of its dual problem.
In addition, $\sf{g}(\bs{\mu})$ is a concave function by nature.
And the sub-gradient for $\mu_m$ is $\dfrac{\bar{Q}_m}{\Delta_T} - \sum \limits_{j = 1}^{\bar{T}_m}  \log_2\left(1+\dfrac{\Gamma_m[t](\mb{w}_t)}{\xi_{\sf{fit}}}\right)$ which can be justified by 
taking $\partial \mathcal{L}(\mb{W},\bs{\mu})/\partial \mu_m$.
Then, the dual variables can be updated iteratively following the sud-gradient method as
\beq \label{mu-update1} 
\mu_m^{[\ell +1]} =\mu_m^{[\ell]} + r_{\ell} \left[\dfrac{\bar{Q}_m}{\Delta_TBW} - \!\! \sum \limits_{t = 1}^{\bar{T}_m} \!\!  \log_2\left(\!1\!+\!\dfrac{\Gamma_m[t](\mb{w}_t^{[\ell]})}{\xi_{\sf{fit}}}\right)\right], 
\eeq
where the suffix $[\ell]$ represent iteration $\ell$ of the inner applied for solving problem \eqref{MinPro_rlxed2}, $\big\{ \mb{w}_t^{[\ell])}\big\} $ are the beamforming vectors at time-slot $t$ interation $\ell$, $r_{\ell}$ is suitable small step-sizes.
If $r_{\ell} \overset{\ell \rightarrow \infty} \longrightarrow 0$,
the above sub-gradient method is guaranteed to converge to the optimal solution of problem \eqref{dual-prob}.
\subsubsection{MMSE-based LP Design}
In what follows, one focuses on developing the LP design for solving problem \eqref{dual_func}. Along with the way updating the Lagrangian parameter $\bs{\mu}$ given in \eqref{mu-update1}, this LP solution facilitate us propose an algorithm dealing with \eqref{MinPro_rlxed2} and \eqref{MIN_POWER_PROB} as well.
For given $\bs{\mu}$, one is worth noting that problem \eqref{dual-prob} can be decoupled into independent  problems corresponding to $T$ TSs, i.e.,
\beq 
\hspace{-0cm}  \underset{\mb{w}_t}{\min} \hspace{-0.1cm} \sum \limits_{\forall n} \! \beta_n^{(k)}\![t] P_n\![t] \! - \!\!\!\!\!\!\! \sum_{m \in \mathcal{M}[t]} \!\!\!\!\!\! \mu_m  f_{\sf{SN}}\!\left( \Gamma_{m}[t](\mb{w}_{t})\right) 
\text{s.t.$(\tilde{C}2,C3)_t$,} \label{Pro_tslot}
\eeq
where $\mathcal{M}[t] = \left\lbrace m\left| \bar{T}_m \geq t\right.  \right\rbrace$ while the low suffix $(*)_t$ added in constraint notations indicates TS $t$. 
The following theorem aims to relating this non-convex problem to a weighted sum-mean square error (MSE) minimization problem.
\begin{theorem}
	\label{P2_thr2}
	Problem \eqref{Pro_tslot} is equivalent to the following
	weighted sum-MSE and power minimization problem, i.e. two problems have same optimal solutions,
\begin{subequations} \label{WMMSE-prob2}
	\begin{eqnarray}
		\hspace{-1cm} &\underset{\mb{w}_t,\bs{\delta}_t,\bs{\omega}_t \bs{\alpha}_t}{\min} &
		\sum \limits_{\forall n} \beta_n^{(k)}[t] P_n[t] + \sum \limits_{m \in \mathcal{M}[t]} \mu_m \alpha_m[t]  \\
	\hspace{-1cm}&\text{s.t. }&  \hspace{-0.7cm}   \text{constraints $(\tilde{C}2)_t$ and $(C3)_t$,} \\
	\hspace{-1cm}&&  \hspace{-0.7cm} (C7)_t:  \alpha_m[t] \geq  \frac{\omega_m[t] e_m[t] -  \ln \omega_m[t]  -  1}{\ln 2} , \forall m, \\
	\hspace{-1cm}&& \hspace{-0.7cm} (C8)_t:  \alpha_m[t]  \geq - R_L, \forall m,
	\end{eqnarray}
\end{subequations}
	where $e_m[t] = \mathbb{E} \left[ \big|s_m[t] - \delta_{m}[t] \tilde{z}_m[t]\big|^2 \right]$, %$\tilde{z}_m[t]$ is a virtual received signal that 
	$\tilde{z}_m[t] = \frac{1}{\sqrt{\xi_{\sf{fit}}}}\mb{h}^H_{m,m}[t] \mb{w}_m[t] s_{m}[t] + \sum \limits_{j \neq m}^{M} \mb{h}^H_{j,m}[t] \mb{w}_j[t] s_{j}[t] + \eta_m[t]$,
	 $\omega_{m}[t]$ and $\delta_{m}[t]$ represent the MSE weight and the receive coefficient corresponding to $\tilde{z}_m[t]$, respectively.
\end{theorem}
\begin{IEEEproof} 
	The proof is given in Appendix~\ref{prf_P2_thr2}. 
\end{IEEEproof}

It is noted that constraint $(C7)_t$ is not \emph{jointly} convex, but it is convex over each set of variables. Hence, this problem can be solved by alternately optimizing $\bs{\delta}_t$, $\bs{\omega}_t$, and $(\mb{w}_t,\bs{\alpha}_t)$. 
Particularly, for given $(\mb{w}_t,\bs{\alpha}_t)$, $\bs{\delta}_t$, and $\bs{\omega}_t$ can be determined according to the results in Appendix~\ref{prf_P2_thr2} as
\begin{eqnarray} \label{receive2}
	\delta_m^{\star}[t] = \delta_m^{\mr{MMSE}}[t] = (1/\sqrt{\xi_{\sf{fit}}})\Theta_m^{-1}[t]\mb{w}_m^H[t] \mb{h}_{m,m}[t]. 
\end{eqnarray}
 where $\Theta_m[t] =\vert\tilde{z}_m[t]\vert^2$ and the optimum value of $\omega_m^{\star}[t]$ can be expressed as
\beqn \label{omega2}
\omega_m^{\star}[t]  =  e_m^{-1}[t]  =  1  + (1/\xi_{\sf{fit}}) \Theta_m^{-1}[t]\vert\mb{w}_m^H[t] \mb{h}_{m,m}[t]\vert^2.
\eeqn
For given $\bs{\delta}_t$, and $\bs{\omega}_t$, the optimal $\mb{w}_t$ can be obtained by solving the following QCQP problem:
\beq
\hspace{0mm}	\underset{\mb{w}_t,\bs{\alpha}}{\min} 
	\sum \limits_{\forall n} \!\! \! \beta_n^{(k)}\![t] P_n\![t] \! + \!\!\!\!\!\! \sum \limits_{m \in \mathcal{M}[t]} \!\!\!\!\! \mu_m \alpha_m\![t] \text{ s.t. $\!(\tilde{C}2)_t,\!(C3)_t,\!(C7)_t,\!(C8)_t$,} \label{QCQP-probW}
\eeq
Here, $P_n[t] \!\!\! =  \!\!\! \!\sum_{\forall m}\!\! \mb{w}_m^H[t] \mb{E}^n_m \mb{w}_m[t]$ and $e_m[t] \!\!\! = \!\!\! \sigma^2_m|\delta_m[t]|^2+1+$ $\sum_{\forall j} \! |\delta_m[t]|^2 \mb{w}_j^H[t] \mb{U}_{j,m} \mb{w}_j[t] \!\!\! - \!\!\! \frac{2}{\sqrt{\xi_{\sf{fit}}}}\Re\left( \delta_m^{\prime}[t]\mb{w}_m^H[t]\mb{h}_{m,m}[t]\right)$ where $\mb{U}_{j,m} = \mb{h}_{j,m}[t] \mb{h}_{j,m}^H[t]$ if $j \neq m$, $\mb{U}_{m,m} =  \frac{\mb{h}_{m,m}[t] \mb{h}_{m,m}^H[t]}{\xi_{\sf{fit}}}$, and 
$\Re(.)$ denotes the real part. This QCQP problem can be solved by any standard convex optimization solvers or the  Lagrangian duality method \cite{Boyd2009}.
By iteratively updating $ \bs{\mu}$ and $\left\lbrace\mb{w}_t, \bs{\alpha}_t, \bs{\delta}_t, \bs{\omega}_t \right\rbrace$'s for all time-slots, we can solve problem \eqref{MinPro_rlxed2} and obtain the
MMSE LP vectors corresponding to given value of $\bs{\psi}$. Combined with the compressed sensing-based method, the sparse LP design for payload power minimization is summarized in Alg.~\ref{P2_alg:2}. 
%The convergence of this algorithm is analyzed in the following theorem.
\setlength{\textfloatsep}{6pt}% Remove \textfloatsep
\begin{algorithm}[!t]%\leesize
\footnotesize
	\caption{\textsc{Joint LP and BH Design}}
	\label{P2_alg:2}
	%\algsetup{indent=1.5em}
	\begin{algorithmic}[1]
		\STATE \textbf{Initialize:} \begin{enumerate}[label = {1-\alph*:}]
			\item Choose non-negative values of $\psi_n^{(0)}[t]$'s and $\mu_m^{[0]}$'s.
			\item Select $\mb{W}^{[0]}$ satisfying constraint $(\tilde{C}2)$ and $(C3)$.
			\item Set $k=0$, $\ell=0$.
		\end{enumerate}  
		\REPEAT 
		\REPEAT
		\FOR{$t=1$ to $T$}
		\STATE Calculate $(\delta_m^{[\ell+1]}[t],\omega_m^{[\ell+1]}[t])$'s as in (\ref{receive2}, \ref{omega2}).
		\STATE Determine $\mb{w}_m^{[\ell+1]}[t]$'s by solving problem (\ref{QCQP-probW}) corresponding to $\bs{\psi}^{(k)}$, $\bs{\mu}^{[\ell]}$, $\delta_m^{[\ell+1]}$'s, and $\omega_m^{[\ell+1]}$'s.
		\ENDFOR
		\STATE Update $\bs{\mu}^{[\ell+1]}$ based on $\bs{\mu}^{[\ell]}$ and $\mb{W}^{[\ell+1]}$.
		\STATE Set $\ell:=\ell+1$.
		\UNTIL Solution of problem \eqref{MinPro_rlxed2} converges.
		\STATE Update $\bs{\psi}^{(k+1)}$'s based on $\mb{W}^{[\ell]}$ solution.
		\STATE Set $k:=k+1$.
		\UNTIL Convergence.
	\end{algorithmic}
	\normalsize
\end{algorithm}

	\vspace{-0.1cm}

\subsection{Solution Return}
\label{sec:Opt_retu}
Assume that Alg.~\ref{P2_alg:2} returns a feasible solution $\mb{W}^{*}$ and $\mb{g}^*$ can be obtained as $\Gamma_m[t](\mb{w}^{*}_t)$ which may not in $\Omega$.
For discrete-solution return, $g_m^*[t]$'s can be  appropriately rounded to their closest values in $\Omega$. Once the 
discrete values $\mb{g}$ satisfying $(C5)$ are obtained, the LP and BH can be optimized again as follows. 
For given $\mb{g}$, problem (\ref{MIN_POWER_PROB}) can be further decomposed into $T$ per-slot problem (PSP) corresponding to $T$ TSs as
\beq
\underset{\mb{W}}{\min} \!\! \sum \limits_{\forall n} \!\! \left( P_n[t] \!+\! \rho_{\sf{hw}} \left\| P_n[t]\right\|_0\right) \text{ s.t. $(C1)_t$, $(C2)_t$, $(C3)_t$.}  \label{SPARSE_POWER_PROB}
\eeq
Employing the re-weighted $\ell_1$-norm minimization method and properly choosing and updating $\psi_n^{(k)}[t]$'s as in the previous section, PSP \eqref{SPARSE_POWER_PROB} can be approximated to the following.
	\begin{subequations} \label{LLP_rlxed}
		\begin{eqnarray} 
			\hspace{-0.8cm}&\underset{\mb{W}}{\min}& \hspace{-0.2cm} \sum \limits_{\forall n}  \left( 1 + \rho_{\sf{hw}} \psi_n^{(k)}[t]\right) \sum_{m \in \mathcal{M}_n} \mb{w}_{m}^H[t] \mb{E}^n_m \mb{w}_{m}[t]  \label{obj_PLL_rlxed}\\
			\hspace{-0.8cm}&\text{s. t. }&  \hspace{-0.2cm}   \sum \limits_{m \in \mathcal{M}_n} \mb{w}_{m}^H[t] \mb{E}^n_m \mb{w}_{m}[t]  \leq \bar{P}^{\mr{GEO}}_{n}, \quad \forall n \label{cnt1a} \\
			\hspace{-0.8cm}&& \hspace{-0.2cm} \Gamma_m[t] \geq   g_m[t], \quad \forall m, \label{cnt2a} \\
			\hspace{-0.8cm}&& \hspace{-0.2cm} \sum \limits_{\forall n} \psi_n^{(k)}[t] \sum \limits_{m \in \mathcal{M}_n} \mb{w}_{m}^H[t] \mb{E}^n_m \mb{w}_{m}[t]  \leq K_t, \label{cnt3a}
		\end{eqnarray}
	\end{subequations}
	This problem is a traditional power minimization precoding design problem which can be solved effectively by employing the SDP method \cite{Bengtsson99}.
	Then, the compress-sensing based approach for solving PSP \eqref{SPARSE_POWER_PROB} is summarized in Alg. ~\ref{P2_alg:1}.
	If problem \eqref{SPARSE_POWER_PROB} is infeasible, the rounded $g_m[t]$ can be adjusted until all PSPs are feasible.
	
	\begin{algorithm}[!t]%\leesize
	\footnotesize
		\caption{\textsc{Solving Per-Slot Problem Algorithm}}
		\label{P2_alg:1}
		%\algsetup{indent=1.5em}
		\begin{algorithmic}[1]
			%\REQUIRE Maximum iteration number $N$, tolerance $\epsilon$.
			\STATE \textbf{Initialize:}  Set $\psi_n^{(0)}[t]=1$ for all $n \in \mathcal{N}$ and $k=0$.
			\REPEAT 
			\STATE Solve problem \eqref{LLP_rlxed} with $\psi_n^{(k)}[t]$'s to achieve $\mb{W}^{(k)}$.
			\STATE Update $\psi_n^{(k+1)}[t]$'s based on $\mb{W}^{(k)}$ as in \eqref{weight_factor3}.
			\STATE Set $k:=k+1$.
			\UNTIL Convergence.
		\end{algorithmic}
	\normalsize
	\end{algorithm}
	
	\vspace{-0.1cm}
	
\section{Heuristic and DNN-based Solutions}
Note that implementing the window-based Alg.~\ref{P2_alg:2} is very complicated when $T$ is sufficiently large. 
Moreover, this method also requires the all CSI of all TSs estimated at the beginning of time window which is very challenging in practical scenarios, e.g., these CSI might be out-dated.  Hence, this section aims to present two methods to define suitable $\mb{g}$. Once $\mb{g}$ is defined, $\mb{w}[t]$'s can be optimized by solving $T$ PSPs as presented in Section~\ref{sec:Opt_retu}.

	\vspace{-0.1cm}

\subsection{Heuristic Solution}
For heuristic approach, $\mb{g}$ is defined so that the MODCOD rate of each user is the same over its transmission period.
Particularly, the $g_m[t]$ of user $m$ over period $[1,\bar{T}_m]$ as follows,
\beq
\hspace{-2mm} g_m[t]=\bar{\gamma}_{l^{\star}_1} | l^{\star}_1= \underset{l}{\argmin} R_l \! \geq \! \lceil \bar{Q}_m/(\Delta_T BW \bar{T}_m) \rceil, t \leq \bar{T}_m,
\eeq
and $g_m[t] = 0$ if $t > \bar{T}_m$.
%g_m[\bar{T}_m] = \bar{\gamma}_{l^{\star}_2} | l^{\star}_2= \underset{l}{\argmin} R_l \geq \leftceil \bar{Q}_m-\Delta_T BW R_{l^{\star}_1}(\bar{T}_m-1) \rightceil.
	\vspace{-0.1cm}

\subsection{DNN-based Solution}
Herein, the DNN method is employed to predict the suitable $\mb{g}$ in time-slot $t$.
The inputs of DNN include the features related to the CSI of the current TS and the remaining anmount of demand data. Particularly, the input of the DNN is given by
\begin{equation}
	X_i^{\sf input}[t] = \left[\left[\mb{h}_{m}^{\sf best}[t], S_m[t]\right]_{m=1:M}, \bs{\tilde{g}}_i[t] \right],
	\label{eq32}
\end{equation}
where $\mb{h}_{m}^{\sf best}[t]$ includes top-$3$ elements of $|\mb{h}_{m}[t]|$, 
$S_m[t]= (\bar{Q}_m - \sum_{j=1}^{t} R_m[j])/(\bar{T}-t+1)$, while $\bs{\tilde{g}}_i[t]$ is a potential action. In the training, ones assumes there are $I$ labels in TS $t$, then there are $I$ inputs $X_i^{\sf input}[t], i=1:I$ corresponding to $I$ potential actions. 
These actions are chosen randomly following the uniform distribution where its mean is the action $\mb{g}[t-1]$. 
Particularly, element ${\tilde{g}}_{i,m}[t]$ is determined as
\begin{equation}
	\bs{\tilde{g}}_{i,j}[t] \sim \mathcal{U}_\Omega(\bar{\gamma}_{\max(0,l - L_0)}, \bar{\gamma}_{\min(L,l + L_0)})
\end{equation}
where $\bar{\gamma}_l = g_j[t-1]$, and $\mathcal{U}_\Omega(l_1, l_2)$ is the projection of the uniform distribution from $l_1$ to $l_2$ on the set $\Omega$.
To get the label for training DNN, we define a metric evaluating the objective and penalty as follows
\begin{equation}
	X^{\sf output}[t] = O[t] + \beta_1 Y_1[t] + \beta_2 Y_2[t]  + \beta_3 Y_3[t]
\end{equation}
where $\beta_1, \beta_2, \beta_3$ are controlled weights, $O[t] = \sum_{\forall (n)}  \left( P_n[t] + \right.$ $\left. \rho_{\sf{hw}} \left\| P_n[t]\right\|_0\right)$, $Y_1[t] \!\! = \!\! \left[\!\frac{1}{M} \!\! \sum_m \!\! \left(g_m[t] -  \Gamma_m[t](\mb{w}_t) \right)  \right]^+$, $Y_2[t]$ $=\left[ \sum_{n \in \mathcal{N}} \left\| P_n[t] \right\|_0 - K_t \right]^+$, $Y_3[t] = \left[ \frac{1}{M} \sum_m \left(R_m[t] - S_m[t] \right) \right]^+$. 
Denote the output of the DNN as $\Phi(X_i^{\sf input}[t])$, then the action in the testing can be given by $\bs{g}^{\sf{out}}[t] = \argmin_i \Phi(X_i^{\sf input}[t])$.
In this paper, the DNN $\Phi(.)$ includes 4 hidden layers, each having 100 neurons, and a Dropout layer after the first hidden layer to prevent the over-fitting problem. The mean-square-error is applied in the loss function, and the rectified linear activation function (ReLU) is applied for all layers. %The detail structure of DNN is shown in Fig.~\ref{fig-DNN-structure}

	\section{Simulation Results}
	\subsection{Channel Model and Data Generation}
	\subsubsection{Channel Model}
	In this simulation, the downlink channel coefficient from antenna of beam $n$ to user $m$, $h^n_m[t]$ is modeled based on Rician model as,
	\beqn
	\hspace{-0.5cm} h_{n,m}[t] \!\!\! & = & \!\!\! e^{-j\left(\frac{2\pi d(\theta^{\sf{la}}_m, \theta^{\sf{lo}}_m)}{\lambda}+\phi^n_m[t]\right)} \left[ G^{\sf{gu}}_m / P_{\sf{loss}} \left(\theta^{\sf{la}}_m, \theta^{\sf{lo}}_m \right)\right]^{1/2} \times    \nonumber \\
	\hspace{-0.5cm}&& \hspace{-0.1cm}    \left[ \sqrt{{L}/{(L+1)}}b^{\sf{pa}}_n\left(\theta^{\sf{la}}_m, \theta^{\sf{lo}}_m \right) +  \sqrt{{1}/{(L+1)}}\alpha^n_m[t] \right] , 
	\eeqn
	where $G^{\sf{gu}}_m$ is the receiving antenna gain; 
	$P_{\sf{loss}} \left(\theta^{\sf{la}}_m, \theta^{\sf{lo}}_m\right)=\left[ {\lambda}/{4 \pi d(\theta^{\sf{la}}_m, \theta^{\sf{lo}}_m)}\right]^2$ is the path-loss, $\theta^{\sf{la}}_m$  and  $\theta^{\sf{lo}}_m$ are latitude and longitude of user $m$; 
	$b^{\sf{pa}}_n\left(\theta^{\sf{la}}_m, \theta^{\sf{lo}}_m \right)$ represents the pattern coefficient of beam $n$  corresponding to user $m$'s location; $\alpha^n_m[t]$ is the small NLoS fading; $L$ denotes LoS/NLoS Rician factor; $d(\theta^{\sf{la}}_m, \theta^{\sf{lo}}_m)$ is the distance between satellite and user $m$; $\lambda$ is the wave length, and $\phi^n_m[t]$ stands for the phase noise which is model as the summation of the phase noises at satellite and user $m$ as $\phi^n_m[t] = \phi^{\sf{GEO}}[t] + \phi^{\sf{gu}}_m[t]$.
	Here, phase noise is one of the imperfections from the hardware components, e.g., oscillators. Then, one assumes that $\phi^{\sf{GEO}}[t]$ is the same for all beams while $\phi^{\sf{gu}}_m[t]$'s vary independently.   
	
	\subsubsection{Channel Data Generation}
	To generate the channel data for one realization, the random-walk process is employed as $
	\phi^{\sf{GEO}}[t+1]  =  (1-\zeta)\phi^{\sf{GEO}}[t] + \zeta\delta^{\sf{GEO}}[t]$ and
	$\phi^{\sf{gu}}_m[t+1] = (1-\zeta)\phi^{\sf{gu}}_m[t] + \zeta\delta^{\sf{gu}}_m[t]$, $\alpha^n_m[t+1] = (1-\zeta)\alpha^n_m[t] + \zeta \xi^n_m[t]$
	where $\zeta$ is random-walk factor, $\phi^{\sf{GEO}}[0]$, $\phi^{\sf{gu}}_m[0]$'s, $\delta^{\sf{GEO}}[t]$, and $\delta^{\sf{gu}}_m[t]$'s are all zero-mean random samples but with difference variances and $\alpha^n_m[0]$ and are semi-static frequency-flat and uncorrelated complex Gaussian random variables with zero mean and unit variance. Note that, $\phi^{\sf{GEO}}[0]$, $\phi^{\sf{gu}}_m[0]$'s, and $\alpha^n_m[0]$'s are generated independently over different realizations.
	
	%	\vspace{-0.3cm}
	
	\subsection{Numerical Results}
			\begin{table}[!t]
			\centering
		\caption{Simulation Parameters}
		\label{tab:simpara}
		\vspace{-0.2cm}
		\begin{tabular}{l | r}
			\toprule
			\midrule
			Satellite Orbit													&$13^{\circ}$E (GEO)		\\
			GEO's Maximum Tx-Power per beam										& $50-100$ W\\
			Beam Hardware-Power (LNA, HPA, RF Conv.)										& $5$ W\\
			Number of Virtual Beams ($N$)					& 10\\	
			Beam Radiation Pattern 					&Provided by ESA \\
			Downlink Carrier Frequency							& 19.5 GHz\\
			User Link Bandwidth, $B$							& 500 MHz\\ 
			Noise Power												& $-118.42$ dB\\
			Number of TSs ($M$)							& 10 \\
			Time-slot duration ($\Delta_T$) & $20$ ms \\
			\bottomrule
		\end{tabular}
		\vspace{-0.2cm}
		\end{table}
	\begin{figure}[!t]
		\centering
		\includegraphics[width=75mm]{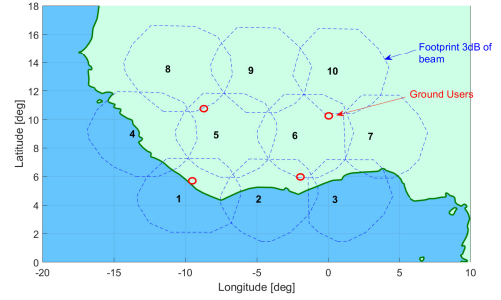}
		\vspace{-0.2cm}
		\caption{Considered GEO multibeam footprint pattern with $N=10$.}
		\label{map_footprint}
			\vspace{-0.5cm}
	\end{figure}

	\begin{figure}[!t]
		\centering
		\includegraphics[width=75mm]{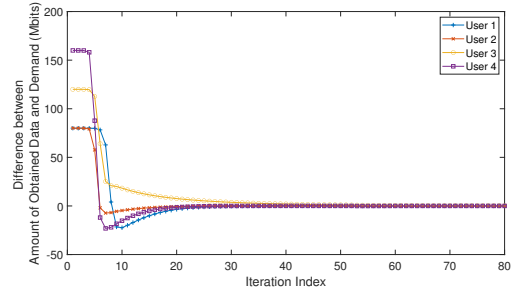}
		\vspace{-0.2cm}
		\caption{Convergence of proposed algorithm.}
		\label{convergence_alg1}
			\vspace{-0.2cm}
	\end{figure}
	
	\begin{figure}[!t]
		\centering
		\includegraphics[width=70mm]{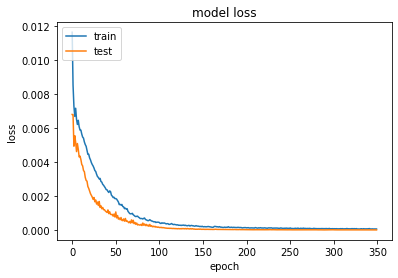}
		\vspace{-0.2cm}
		\caption{DNN model training loss versus the number of epochs.}
		\label{convergence_DNN}
			\vspace{-0.2cm}
	\end{figure}
	We consider a GEO satellite system with $10$ spot beams, i.e., $N=10$ as shown in Fig.~\ref{map_footprint}. The setting parameters are summarized in Table \ref{tab:simpara}.
	Four users are considered in this simulation where $\bar{Q}_m$'s and $\bar{T}_m$'s are set at $[200,200,300,400]$ (Mbits) and $[7,5,6,10]$, respectively. Firstly, we examine the convergence of Alg.~\ref{P2_alg:2} where the difference between the obtained rate and the demand due to four users is illustrated over iterations in Fig.~\ref{convergence_alg1}. As can be seen, the differences vary before converge at zeros after around $50-60$ iterations. In addition, Fig.~\ref{convergence_DNN} is presented to the loss function of DNN model versus the number of training epochs. This figure has confirmed the convergence of the proposed DNN-based solution approach. 

	\begin{figure}[!t]
		\centering
		\includegraphics[width=75mm]{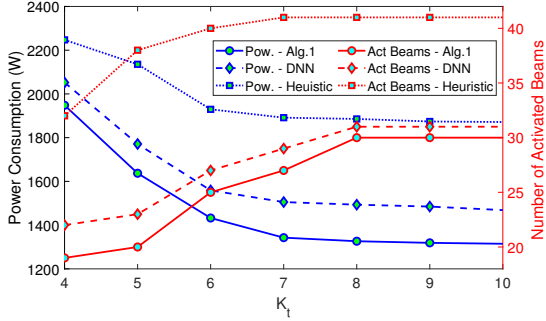}
		\vspace{-0.2cm}
		\caption{Power versus $K_t$.}
		\label{pow_Kt}
			\vspace{-0.5cm}
	\end{figure}
	
		\begin{figure}[!t]
		\centering
		\includegraphics[width=75mm]{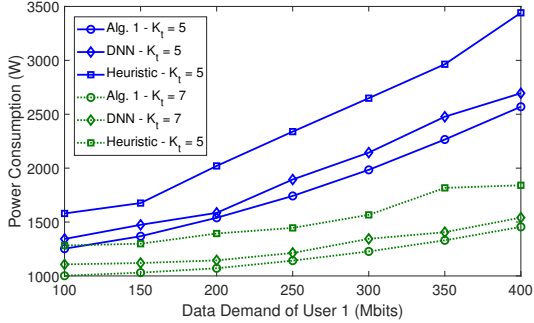}
		\vspace{-0.2cm}
		\caption{Power versus $\bar{Q}_1$.}
		\label{pow_Q1}
			\vspace{-0.2cm}
	\end{figure}
Fig.~\ref{pow_Kt} presents the variations of the payload consumed power and the total number of activated beams according three proposed algorithms versus $K_t$ while Fig.~\ref{pow_Q1} illustrates the consumed-power variation when the required amount of data of user $1$ is varied from $100$ to $400$ (Mbits) while that of these others are unchanged.
As expected, the consumed power due to all three solution approaches decreases as $K_t$ increases and increases if user $1$ requires more data traffic. 
In addition, Alg.~\ref{P2_alg:2} can provide the lowest payload power as well as activate the smallest number of beams in all TSs to meet the user's demands.
The DNN-based method also outperforms the heuristic one significant and achieve the outcomes closed to Alg.~\ref{P2_alg:2}.
This is quite impressive since DNN-based (and heuristic) implementation does not need all CSI according to all TSs at the beginning of the time window. 

\vspace{-0.2cm}

	\section{Conclusion} \label{ccls}
	This paper has considered new joint LP and BH designs for multi-beam GEO satellite communication systems. We have proposed one window-based optimizing algorithm and two two-phase solutions, using heuristic approach or DNN tool, to determined the sparse LP, BH, and MODCOD selection to minimize the power consumed by the payload and meet various data demands from user terminals across the coverage area.
	Numerical results have confirmed the superior performances of the window-based algorithm as well as the DNN-based solution approach.
	
	\appendices
	
	\section{Proof of Theorem~\ref{P2_thr2} }
	\label{prf_P2_thr2}
	Employing $\delta_m[t]$, the estimated symbol can be given by $\hat{s}_m[t]=\delta_m[t]\tilde{z}_m[t]$. 
	Based on the MMSE-receiving filter, the receiving coefficient can be optimized as 
	$\delta_m^{\star}[t] = \arg\min_{\delta} \mathbb{E}\big\{\left|s_m[t] - \delta \tilde{z}_m[t]\right|^2\big\}$. The result of RHS is given in \eqref{receive2}. 
	Then, the corresponding MSE for user $m$ can be described as
	$ e_m[t] = \mathbb{E}\big\{\left|s_m[t] - \delta_m^{\mr{MMSE}}[t] \tilde{z}_m[t]\right|^2\big\} = \big(1 + \frac{1}{\xi_{\sf{fit}}} \Gamma_m\left(\mb{w}_t\right) \big)^{-1}$.
	Hence, $f_{\sf{SN}}\left( \Gamma_{m}[t](\mb{w}_{t})\right)$ can be expressed as a function of
	$e_m[t]$ as 
	$f_{\sf{SN}}\left( \Gamma_{m}[t](\mb{w}_{t})\right) = \min\left[R_L, \log_2(1/e_m[t]) \right]$.
	Furthermore, employing the first-order Taylor approximation for the $\log$-function yields
	$\ln(1/e_m[t]) \geq \ln(\omega_m[t])+\omega_m[t](\omega_m^{-1}[t] - e_m[t])$ whose equality holds at $\omega_m^{\star}[t] = e_m^{-1}[t] $.
	
	In addition, at $\delta_m^{\star}[t]$ and $\omega_m^{\star}[t]$, 
	one has $f_{\sf{SN}}\left( \Gamma_{m}[t](\mb{w}_{t})\right) = -\max \left(-R_L, k_m[t] \right)$ where $k_m[t]=\frac{\omega_m[t]e_m[t] - \ln(\omega_m[t])-1}{\ln 2}$. 
	Hence, problem \eqref{Pro_tslot} is equivalent to 
		\begin{eqnarray} 
		\hspace{-0.4cm} &\underset{\mb{w}_t,\bs{\delta}_t, \bs{\omega}_t}{\min}& \hspace{-0.2cm} \sum \limits_{\forall n} \beta_n^{(k)}[t] P_n[t] +  \sum_{m \in \mathcal{M}[t]} \mu_m \max \left(-R_L, k_m[t] \right) \nonumber\\
		\hspace{-0.4cm}&\text{s. t. }&  \hspace{-0.2cm}   \text{constraints $(\tilde{C}2)_t$ and $(C3)_t$,} \label{Pro_tslot2}
	\end{eqnarray}
	since they share the same solution of $\mb{w}_t$ at the optimum points \cite{VuHa_TGCN20}.
%Enhancing these results with the help of some minor manipulations, one can show that problem \eqref{Pro_tslot} and \eqref{WMMSE-prob2} are equivalent since they share the same solution of $\mb{w}_t$ at the optimum points \cite{VuHa_TGCN20}.
	Problem \eqref{Pro_tslot2} then can be rewritten as problem \eqref{WMMSE-prob2} by using the additional variable $\bs{\alpha}_t$.
	The proof thus follows.
	
	\section*{Acknowledgment}
	This work has been supported in parts by the Luxembourg National Research Fund (FNR) under the projects FlexSAT (C19/IS/13696663) and ARMMONY (FNR16352790).
	
	\bibliographystyle{IEEEtran}

\begin{thebibliography}{19}
	    \bibitem{Sat_Survey_21}
	    O. Kodheli, et. al.,``Satellite communications in the new space era: A survey and future challenges'' {\em IEEE Commun. Surveys Tuts.,} vol. 23, no. 1, pp. 70--109, 2021.
        
        \bibitem{Vazquez_WC16}
        M. A. Vazquez, et. al., ``Precoding in multibeam satellite communications: Present and future challenges,'' {\em IEEE Wireless Communications,} vol. 23, no. 6, pp. 88--95, December 2016.
        
         \bibitem{Ginesi_WiSATS17}
        A. Ginesi, E. Re, and P. Arapoglou, ``Joint beam hopping and precoding in HTS systems,'' in {\em 9th Int. Conf. on Wireless and Satellite Systems (WiSATS),} 2017.
        
        \bibitem{ESA_FlexPreDem}
        ESA Project FlexPreDem, ``Demonstrator of precoding techniques for flexible broadband satellite systems,'' https://artes.30esa.int/projects/flexpredem, 2020.
        
        \bibitem{Kibria_Globe19}
        M. G. Kibria, E. Lagunas, N. Maturo, D. Spano, and S. Chatzinotas, ``Precoded cluster hopping in multi-beam high throughput satellite systems,'' in {\em 2019 IEEE GLOBECOM}, pp. 1--6, 2019.
        
        \bibitem{Eva_Frontier21}
        E. Lagunas, et. al., “Precoded cluster hopping for multibeam GEO satellite communication systems,” {\em Frontiers in Signal Processing,} 2021. 
        
		\bibitem{ETSI_DVBS2X}
		{\em ETSI EN 302 307-2 V1.1.1: Second generation framing structure, channel coding and modulation systems for Broadcasting, Interactive Services, News Gathering and other broadband satellite applications}, Part 2: DVB-S2 Extension (DVB-S2X) (2014-10).
		
		\bibitem{Candes08}
		E. Candes, M. Wakin, and S. Boyd, ``Enhancing sparsity by reweighted $\ell_1$ minimization,'' {\em J. Fourier Analysis Applications,} vol. 14, no. 5, pp. 877--905, Dec. 2008.
	
		\bibitem{Boyd2009}
		M.~Grant and S.~Boyd, ``CVX: Matlab software for disciplined convex
			programming (web page and software), 2009. %[Online]. Available:
			%http://stanford.edu/\~{}boyd/cvx},'' June 2009.
		
		
		\bibitem{Bengtsson99}
		M.~Bengtsson and B.~Ottersten, ``Optimal downlink beamforming using
		semidefinite optimization,'' {\em in Proc. Annual Allerton Conf. on Commun.},
		pp.~987--996, Sept. 1999.
		
		\bibitem{VuHa_TGCN20}
		V. N. Ha, D. H. N. Nguyen and J. -F. Frigon, ``System energy-efficient hybrid beamforming for mmWave multi-user systems'', {\em IEEE Trans. on Green Commun. and Net.}, vol. 4, no. 4, pp. 1010--1023, Dec. 2020.
		
		
	\end{thebibliography}

\end{document}